\documentclass[aps,prb,twocolumn,superscriptaddress,showpacs,amsmath,floatfix,citeautoscript]{revtex4-2}

\usepackage{amssymb}
\usepackage{amsmath}
\usepackage{amsthm}
\usepackage{color}
\usepackage{graphicx}

\begin{document}
\title{First-principles calculations of the surface states of doped and alloyed topological materials via band unfolding method}
\author{Zujian Dai}
\affiliation{CAS Key Laboratory of Quantum Information, University of Science and Technology of China, Hefei 230026, Anhui, China}
\affiliation{Synergetic Innovation Center of Quantum Information and Quantum Physics, University of Science and Technology of China, Hefei, 230026, China}
\author{Gan Jin}
\affiliation{CAS Key Laboratory of Quantum Information, University of Science and Technology of China, Hefei 230026, Anhui, China}
\affiliation{Synergetic Innovation Center of Quantum Information and Quantum Physics, University of Science and Technology of China, Hefei, 230026, China}
\author{Lixin He}%
\email{helx@ustc.edu.cn}
\affiliation{CAS Key Laboratory of Quantum Information, University of Science and Technology of China, Hefei 230026, Anhui, China}
\affiliation{Synergetic Innovation Center of Quantum Information and Quantum Physics, University of Science and Technology of China, Hefei, 230026, China}


\begin{abstract}

One of the most remarkable characteristics of topological materials is that they have
special surface states, which are determined by the topological properties of their bulk materials.
The angle resolved photoemission spectroscopy (ARPES)  is a powerful tool to explore the surface states, which allows to further investigate the
topological phase transitions.
However, it is very difficult to compare the first-principle calculated band structures to the ARPES results,
when the systems are doped or alloyed, because the band structures are heavily folded.
We develop an efficient band unfolding method based on numerical atomic orbitals (NAOs). We apply this method to study the
surface states of the non-magnetically and magnetically doped topological insulators Bi$_2$Se$_3$ and the topological crystalline insulators Pb$_{1-x}$Sn$_{x}$Te.

\end{abstract}


\maketitle


\section{introduction}

One of the most remarkable characteristics of topological materials is that they have
special surface states, which are determined by the topological properties of their bulk materials.
For example, a topological insulator (TI) \cite{Hasan&Kane2010,Qi&Zhang2011,Ando2013,Bernevig+2013}
has a gapless surface state, whereas a Weyl semimetal has Fermi arc on its surface\cite{Wan2011,Fangzhong2012,Hasan2015,WengHM2015}.
The gapless states on the surface of TIs are protected by the time reversal symmetry.
When doped by the magnetic impurities, a gap would appear in the surface state, which indicated the topological
phase transition. In these senses,  angle resolved photoemission spectroscopy (ARPES)\cite{Yang2018,Shen2021}
is a powerful tool to investigate the topological properties of a material.

Even though, the topological phase transitions in many materials are now
clear theoretically, it is still very beneficial to be able to compare the energy band structures from
first-principles calculations directly with the ARPES results, which
is however very challenging,  especially for the doped, alloyed and disordered systems.
These calculations usually need to construct very large supercell, which on one hand computational costly,
and on the other hand since the band structure of the supercell is heavily folded in the first Brillouin zone (BZ),
it can not be compared with the ARPES experiments directly\cite{Belita1997,Zunger1998,Kuwei2010}.

  In many cases, the doped and alloyed systems can be viewed as perturbations to the original crystal structure, which break the
  translation symmetry of the original unit cell, and introduce the coupling between
  the different $\boldsymbol{k}$ points in the BZ. We may project the wave functions of the supercell to the coupled $\boldsymbol{k}$ points
  in the original unit cell, to obtain the spectral function, which is known as the band unfolding method
  \cite{Belita1997,Zunger1998,Kuwei2010}.
  The unfolded spectra can then be directly compared with the ARPES experiments.

  The band unfolding method has been implemented in various ways,
  and despite the successful applications of these methods \cite{Kuwei2010,Zunger2012,Lee_2013,Vanderbilt2013,Haiping2013,Jonas2014,Chenmx2018},
  they also have some difficulties in treating large complicated systems.
The band unfolding methods were implemented on the plane wave (PW) bases \cite{Zunger2012,Chenmx2018},
which is very computational demanding.
 Ku et al. unfolded the bands utilizing localized Wannier functions (WFs)\cite{Kuwei2010}.
 However, for large systems the construction of the WFs can be time-consuming,
 and what is worse is that it is difficult to find the one to one correspondence
 between the WFs defined in the supercell and projected cell (PC) when the perturbation is strong.
 Lee and co-authors developed a band unfolding method based on numerical atomic orbitals (NAOs)\cite{Lee_2013}, which
 partially solved the problems of WFs.
 However, it still needs to assume the structure and the corresponding NAO bases set of the PC, and therefore,
 may also not suitable for the systems with strong perturbations, where the atomic positions and chemical compositions
 of the supercells are quite different from the PC.

In this work, we develop an efficient  band unfolding method based on NAO bases.
Unlike the previous method using NAO bases~\cite{Lee_2013}, we expand the NAO Bloch wave functions on the PW bases of the PC,
and therefore, do not need to assume the crystal structures of PC.
The method is transparent, and can be applied to study systems of strong perturbations.
We use this tool to investigate the topological phase transitions of non-magnetically and magnetically
doped topological insulator Bi$_{2}$Se$_{3}$ and the topological crystalline insulator Pb$_{1-x}$Sn$_{x}$Te.

\section{band unfolding in NAO bases}

\begin{figure}
	\centering
	\includegraphics[width=0.45\textwidth]{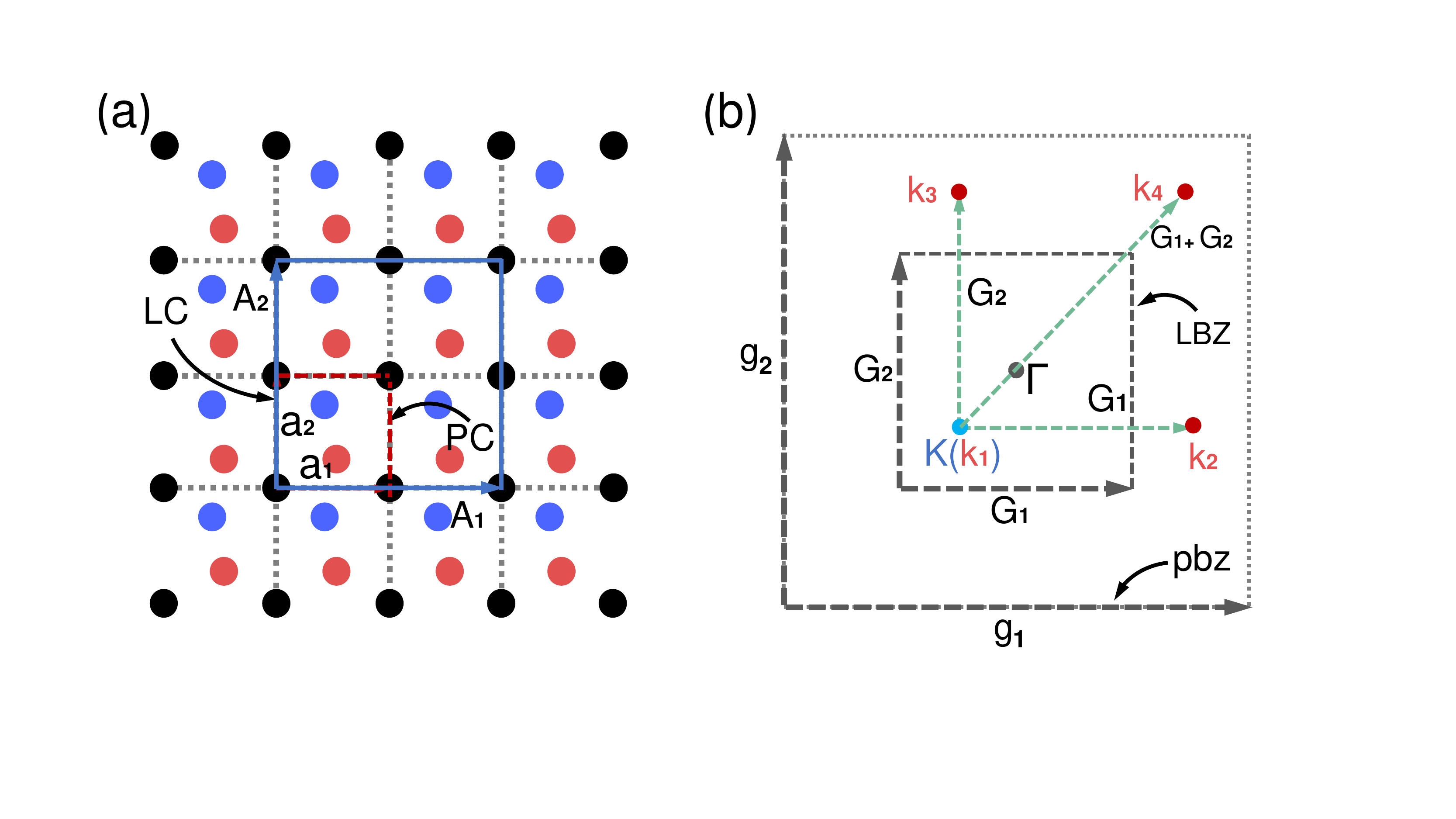}
	\caption{Illustration of (a) the relation between the large cell (LC) and projected cell (PC), and (b) the corresponding BZs.
$\boldsymbol{k}_1$ - $\boldsymbol{k}_4$  of the PC are folded to $\boldsymbol{K}$ point of the LC.
LBZ and pbz are the first BZs of the LC and PC, respectively.}
	\label{fig:LC-PC}
\end{figure}

For convenience, we name the supercell to study, the large cell (LC), which is deformed from the periodically repeated PC.
Figure \ref{fig:LC-PC}(a) schematically shows the relation between the LC and PC,
which is often chosen to be the primitive cell of some structures.
We use the capital letters for the LC, and lower-case letters for the PC.
 The relationships between the lattice vectors of the LC ($\boldsymbol{A}$) and PC ($\boldsymbol{a}$) are given by,
\begin{equation}
\left(
\begin{matrix}
\boldsymbol{A}_1\\
\boldsymbol{A}_2\\
\boldsymbol{A}_3\\
\end{matrix}
\right)=
\left( \begin{matrix}
m_{11} & m_{12} & m_{13}\\
m_{21} & m_{22} & m_{23}\\
m_{31} & m_{32} & m_{33}
\end{matrix} \right) \cdot
\left( \begin{matrix}
\boldsymbol{a}_1\\
\boldsymbol{a}_2\\
\boldsymbol{a}_3
\end{matrix} \right), \quad m_{ij} \in Z \, ,
\end{equation}
or in short, $\boldsymbol{A} = \boldsymbol{M} \cdot\boldsymbol{a}$.
The  $m_{ij}$ are usually taken to be integers to ensure that the PC commensurate
with the LC.
The reciprocal lattice vectors have the similar relation,
 $\boldsymbol{g} = \boldsymbol{M}^{T}\cdot\boldsymbol{G}$, where $\boldsymbol{G}$ and $\boldsymbol{g}$ are the reciprocal lattice vectors of the
 LC, and PC, respectively. The $ \boldsymbol{g}$-vectors belong to a subset of $\boldsymbol{G}$-vectors, as shown in Fig.\ref{fig:LC-PC}(b).

The impurities, defects, etc. in the LC break the translation symmetry of the PC, and therefore introduce the coupling between
different $ \boldsymbol{k}$ points in the BZ of the PC.
The wave function of the LC at  $\boldsymbol{K}$ can be written as the supposition of the contributions
from a series of $\boldsymbol{k}$ points in the PC.
i.e.,
\begin{equation}
|\Psi_{\boldsymbol{K}N}\rangle = \sum_{n, \boldsymbol{k}_{p}} |\psi_{\boldsymbol{k}_{p}n}\rangle\langle \psi_{\boldsymbol{k}_{p}n}|\Psi_{\boldsymbol{K}N}\rangle\, ,
\label{eq:projection}
\end{equation}
where $\sum_{n} | \psi_{\boldsymbol{k}_{p}n} \rangle$ are a complete set of Bloch functions at $\boldsymbol{k}_{p}$.
Note that only the $\boldsymbol{k}$ points that satisfy the pseudo-periodicity relationship [see Fig~\ref{fig:LC-PC}(b)] \cite{Belita1997,Zunger1998},
\begin{equation}
\boldsymbol{k}_{p} = \boldsymbol{K} + \boldsymbol{G}_{p}, \qquad p = 1, \cdots , \det (M)
\end{equation}
contribute to the wave function of the $\boldsymbol{K}$ point.

The spectral function at each $\boldsymbol{k}_{p}$ can be express as follows \cite{Belita1997,Zunger1998},
 \begin{equation}
A(\boldsymbol{k}_{p},E)=  \sum_{N,n} |\,D_{N}(\boldsymbol{k}_{p},n)\,|^2 \delta(E_{N}-E)
\end{equation}
where,
\begin{equation}
D_{N}(\boldsymbol{k}_{p},n) = \langle \psi_{n\boldsymbol{k}_{p}}\,|\,\Psi_{N\boldsymbol{K}}\rangle\, .
\end{equation}

In the NAO bases, the Bloch wave functions of the $N$-th band at $\boldsymbol{K}$ point can be expressed as,
\begin{equation}
\Psi_{\boldsymbol{K}N} = \frac{1}{\sqrt{\mathcal{N}}}\,\sum_{\boldsymbol{R}}\sum_{\mu,i}\,C_{N\mu,i}(\boldsymbol{K})\,e^{i\,\boldsymbol{K}\cdot \boldsymbol{R}}\phi_ {\mu}(\boldsymbol{r}-\boldsymbol{\tau}_{\alpha i}-\boldsymbol{R})\, ,
\end{equation}
 where $\phi_{\mu}(\boldsymbol{r}-\boldsymbol{\tau}_{\alpha i}-\boldsymbol{R})$  are the $\mu$-th atomic orbitals centering on the $i$-th atom of type $\alpha$ in the $\boldsymbol{R}$-th LC.
 The orbital index $\mu$ is a compact notation, i.e.,  $\mu=({\alpha, l , m , \zeta })$ , with $l$ being the angular momentum,
 $m$ the magnetic quantum number,  whereas $\zeta$ is the multiplicity of the atomic orbitals for a given $l$.
 $C_{N\mu,i}(\boldsymbol{K})$ are the coefficients of the bases, and $\mathcal{N}$ is the number of LCs in the Born-von-Karmen supercell.

Generally, the projection functions $| \psi_{\boldsymbol{k}_{p}n}\rangle$  can take the form of,
\begin{equation}
\psi_{\boldsymbol{k}_{p}n}(\boldsymbol{r}) ={1 \over \sqrt{\mathcal{N} V}}
\sum_{\boldsymbol{g}} B_{n\boldsymbol{k}_{p}}(\boldsymbol{g})e^{i(\boldsymbol{k}_{p}+\boldsymbol{g})\cdot\boldsymbol{r}}\, ,
\end{equation}
 where $V$ is the volume of the LC.
Since $| \psi_{\boldsymbol{k}_{p}n}\rangle$  is orthonormal, we have
\begin{equation}
\sum_{\boldsymbol{g}} B^{*}_{n\boldsymbol{k}_{p}}(\boldsymbol{g})B_{m\boldsymbol{k}_{p}}(\boldsymbol{g}) = \delta_{nm}\, .
\end{equation}
One way to chose $\psi_{\boldsymbol{k}_{i}n}(\boldsymbol{r})$ is to use the eigen-states  of some energy bands of the PC.
However, in this case, one has to assume the structure of the PC.
If the studied system is somehow very different from the hypothetical PC, e.g., in the case of alloyed structures, one may need a large amount of
eigen-states as the projection functions.

In this work, we take a different strategy that we use the plane waves directly, i.e., $\psi_{\boldsymbol{k}_{p} ,\boldsymbol{g}}(\boldsymbol{r})
={1 \over \sqrt{\mathcal{N} V}} e^{i(\boldsymbol{k}_{p}+\boldsymbol{g})\cdot\boldsymbol{r}}$  as the projector functions, and therefore,
we do not have to assume the PC structure, as one did in Ref.\onlinecite{Lee_2013}.
In this case, Eq. (6) can be evaluated as follows,
\begin{widetext}
\begin{equation}
D_{N}(\boldsymbol{k}_{p},\boldsymbol{g}) =\frac{1}{\mathcal{N}\sqrt{V}} \sum_{\boldsymbol{R}}\sum_{\mu ,i} \int d\boldsymbol{r}\,
e^{-i(\boldsymbol{k}_{ p}+\boldsymbol{g})\cdot\boldsymbol{r}}
C_{N\mu,i}(\boldsymbol{K})\,e^{i\,\boldsymbol{K} \cdot \boldsymbol{R}}\phi_{\mu}(\boldsymbol{r}-\tau_{\alpha i}-\boldsymbol{R}) \, .
\label{eq:DN}
\end{equation}
\end{widetext}
%
The integration over  $d\boldsymbol{r}$ gives us,
\begin{equation}
\begin{aligned}
&\quad\,\,\,{1 \over \sqrt{V}}  \int d\boldsymbol{r}\,\phi_{\mu}(\boldsymbol{r}-\boldsymbol{\tau}_{\alpha i}-\boldsymbol{R}) e^{-i(\boldsymbol{k}_{p}+\boldsymbol{g})\cdot \boldsymbol{r}} \\
&= \phi_{\mu}(\boldsymbol{k}_{p}+\boldsymbol{g})S_{\alpha, i} (\boldsymbol{k}_{p}+\boldsymbol{g})
e^{-i (\boldsymbol{k}_{p}+\boldsymbol{g}) \cdot \boldsymbol{R}}\, ,
\end{aligned}
\label{eq:lcao-pw}
\end{equation}
where,
\begin{eqnarray}
\phi_{\mu}(\boldsymbol{q}) &=& {1 \over \sqrt{V}} \int d\boldsymbol{r} \,\phi_{\mu}(\boldsymbol{r}) e^{-i \boldsymbol{q}\cdot \boldsymbol{r}} \, ,\\
S_{\alpha, i}(\boldsymbol{q}) &= & e^{-i \boldsymbol{q} \cdot \tau_{\alpha, i}} \, .
\end{eqnarray}
$\phi_{\mu}(\boldsymbol{q})$ is known as the form factor of the orbital, which is determined only by the shape of the orbital,
whereas the structure information is enclosed in $S_{\alpha, i} (\boldsymbol{q})$.
The details of the evaluations of $\phi_{\mu}(\boldsymbol{q})$ are given in the Appendix.

Plugging Eq. (\ref{eq:lcao-pw}) into Eq. (\ref{eq:DN}), and using the the relation $e^{-i(\boldsymbol{k}_{p}+\boldsymbol{g}-\boldsymbol{K})\cdot \boldsymbol{R}}=1$,
we obtain,
\begin{equation}
D_{N}(\boldsymbol{k}_{p},\boldsymbol{g})
=\sum_{\mu,i} \,\phi_{\mu}(\boldsymbol{k}_{p}+\boldsymbol{g})S_{\alpha, i} (\boldsymbol{k}_{p}+\boldsymbol{g})  C_{N\mu,i}(\boldsymbol{K})
\label{eq:DN-2}
\end{equation}

The spectral weight and the spectral function can be express as,
\begin{equation}
A(\boldsymbol{k}_{p},E)=  \sum_{N,\boldsymbol{g}} |\,D_{N}(\boldsymbol{k}_{p},\boldsymbol{g})\,|^2 \delta(E_{N}-E)\,.
\label{eq:spectra}
\end{equation}
The above formulism can be easily generalized to the spin-polarized and spinor wave functions.

This band unfolding method is  physically transparent that it directly measures the spectra of the incoming
plane waves of given momentum $\hbar \boldsymbol{k}$. One does not need to assume
the crystal structure of the PC, or use the band structure information of the PC, and it therefore can be applied to the
structures with strong perturbations.

Equations  (\ref{eq:DN})-(\ref{eq:spectra})  can be efficiently calculated.
In fact, the number of $\phi_{\mu}(\boldsymbol{q})$ is limited, which is only related to the types of the elements in the LC,
and also the number of $ \boldsymbol{g}$ vectors is determined by the size of the PC, which is usually (not always) very small.
Furthermore, for the purpose of spectra calculations, the energy cutoff for the $ \boldsymbol{g}$ vectors
can be much lower than those for the self-consistent and band structure calculations.
Combined with efficient {\it ab inito} calculations utilizing NAO bases, the method allows to calculate unfolded spectra of very complicated systems,
e.g., the surface states of the topological materials.

\section{Surface states of topological materials}

We apply the band unfolding method developed in this work to study the ARPES
of the surface states of topological materials.
We choose to study two series materials: topological insulators and topological crystalline insulators.
For the topological insulators, we calculate the non-magnetically Tl-doped  and the magnetically Fe-doped Bi$_2$Se$_3$.
For the topological crystalline insulators, we investigate the Pb$_{1-x}$Sn$_{x}$Te alloys.

\subsection{Computational details}

The first-principle calculation are carried out with the Atomic orbtial Based Ab-initio Computation at UStc (ABACUS) code\cite{Chen_mohan2010,Lipengfei2016}
within the Perdew-Burke-Ernzerhof\cite{Perdew1996} generalized gradient approximation (GGA) for the exchange-correlation functional.
The ABACUS code is developed to perform large-scale density functional theory (DFT) calculations based on the NAO bases\cite{Chen_mohan2010}.
The optimized norm-conserving Vanderbilt (ONCV) \cite{Hamann2013}
fully relativistic  pseudopotentials \cite{Theurich2001} from the PseudoDojo library\cite{Van2018} are used.

For the Fe and Tl doped Bi$_2$Se$_{3}$, the valence electrons for Fe, Bi and Se are
3s$^{2}$3p$^{6}$3d$^{6}$4s$^{2}$, 5d$^{10}$6s$^{2}$6p$^{3}$ and 3d$^{10}$4s$^{2}$4p$^{4}$, respectively, and
 the NAO bases for Fe, Bi and Se are 4s2p2d1f, 2s2p2d and 2s2p2d, respectively.
 The valence electrons of Tl are 5d$^{10}$6s$^{2}$6p$^{1}$ and NAO bases are
 2s2p2d.

For the Pb$_{1-x}$Sn$_{x}$Te alloys, the valence electrons for Pb, Sn and Te are
 5d$^{10}$6s$^{2}$6p$^{2}$, 4d$^{10}$5s$^{2}$5p$^{2}$ and 4d$^{10}$5s$^{2}$5p$^{4}$.
The 2s2p2d NAO bases are used for the Pb, Sn and Te elements.

In the self-consistent and band structure calculations,
the energy cutoff for the wave functions is set to 100 Ry.
The structures are fully optimized until all forces are less than 0.05 eV/\AA.
During the structural relaxations, the DFT-D3 correction is used to  account the van der waals interactions\cite{Grimme2010}.
The spin-orbit coupling is turned off for the structure relaxations, and turned on
in the self-consistent and band structure calculations.
When calculating the unfolded spectra, the energy cutoff is set to 20 Ry, which already converge the results very well.

\subsection{Results and discussion}

\subsubsection{ARPES for Fe, Tl doped Bi$_2$Se$_3$ }
\begin{figure*}
	\centering
	\includegraphics[width=0.6\textwidth]{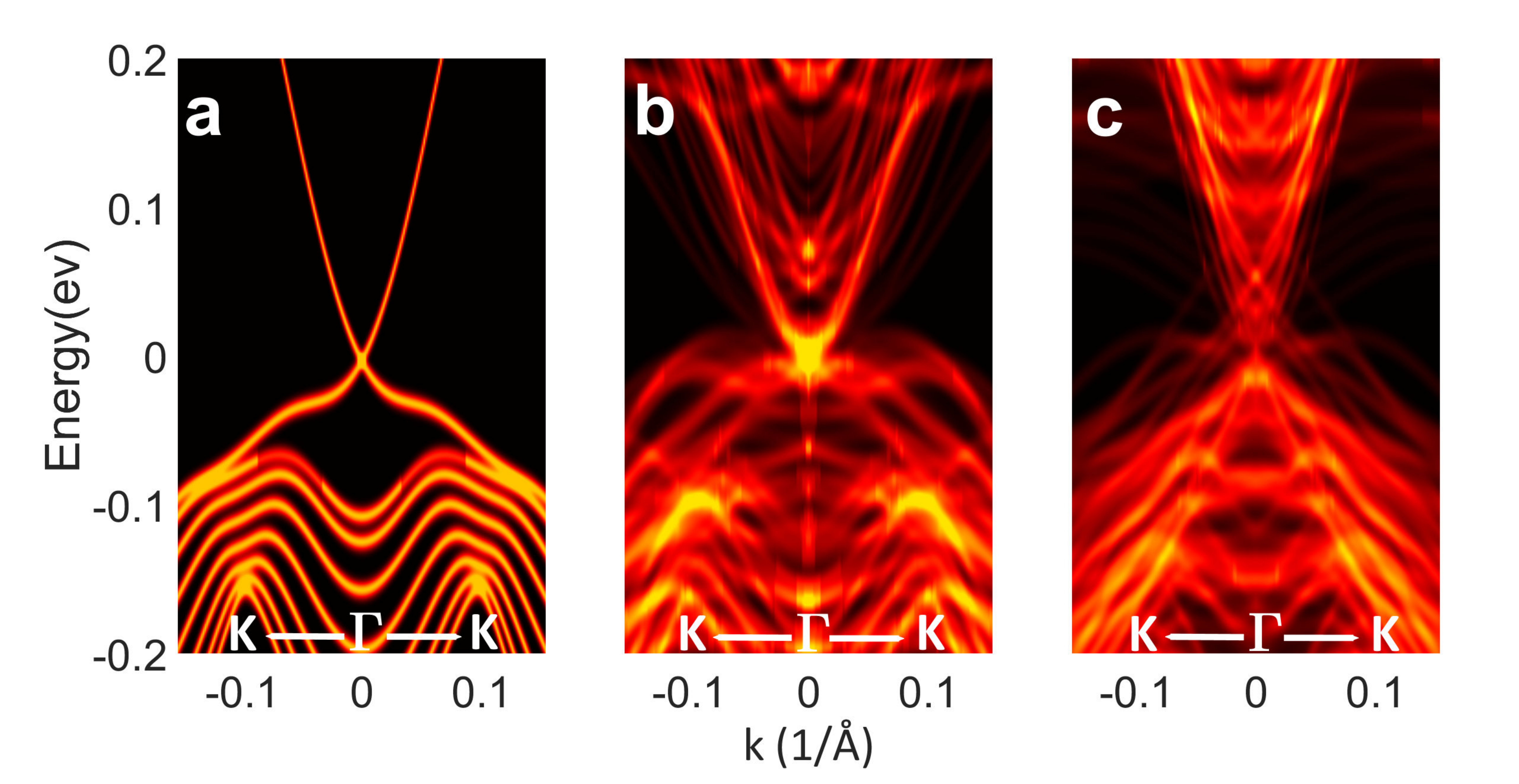}
	\caption{ The ARPES spectra of (a) perfect Bi$_2$Se$_3$ slab;
(b) non-magnetic Tl doped Bi$_2$Se$_3$ slab;  (c) magnetic Fe doped Bi$_2$Se$_3$ slab.
The spectra is broadened by 30 meV }
	\label{fig:Bi2Se3}
\end{figure*}

First-principles band structure calculations \cite{Zhang2009} predicted Bi$_2$Se$_3$, as well as  Bi$_2$Te$_3$  and Sb$_2$Te$_3$
to be 3D strong topological insulators \cite{Fu2007,Moore2007,Roy2009}.
One of the salient  properties of these topological insulators is that they have robust Dirac cone-like, gapless surface band structures, which are
protected by the time-reversal symmetry.
Shortly after the theoretical proposals, the ARPES investigations on Bi$_2$Se$_3$\cite{Hor2009,Hsieh2009_NATURE,Park2010} Bi$_2$Te$_3$\cite{Chen2009,Hsieh2009_PRL,Hsieh2009_SCIENCE} confirmed the theoretical predictions.
Since the gapless surface states are protected by the time reversal symmetry, Dirac nodes should be
robust against the non-magnetic disorder but open gaps in the presence of magnetic impurities\cite{Hsieh2009_NATURE,Xia2009,Chen2009}.
Indeed, the APRES\cite{Chen2010} measurements show that the Tl-doped (Bi$_{0.9}$Tl$_{0.1}$)$_{2}$Se$_3$ does not exhibit a gap opening at the Dirac point, whereas
Fe-doped  (Bi$_{0.84}$Fe$_{0.16}$)$_{2}$Se$_3$ and Mn-doped (Bi$_{0.99}$Mn$_{0.01}$)$_{2}$Se$_3$
show gap openings at the Dirac point, which confirmed
the theoretical predictions.

 Direct first-principle calculations of the surface state of TI is computational costly. \cite{Takehiro2015,Xueqikun_2015,Sanjeev_2020,Shirali_2020}
 In previous calculations, one often adapts $\boldsymbol{k}\cdot\boldsymbol{p}$  methods\cite{Liuchaoxing2010}
 or using the tight-binding models based on WFs\cite{Mostofi2008} constructed from bulk calculations.
 These methods ignore the effects of structure relaxation and chemical environment changes of the surface to the WFs.
 To calculate the surface states of doped or alloyed TI is even more difficult, as much larger supercells are required,
 and the band unfolding through WFs can be very cumbersome\cite{Kuwei2010,Vanderbilt2013}.
  To study the doping effect of the surface of TI, virtual crystal approximation (VCA) with tight-binding methods\cite{Vanderbilt2013}
 has been used, which somehow ignores the fluctuation effects of the structure and chemical compositions.

To study the surface states of the Fe and Tl doped  Bi$_2$Se$_3$, we construct a slab containing 2$\times$2$\times$9 Bi$_2$Se$_3$  five-atom unit cells.
A 15 \AA ~vacuum is added to avoid the interactions between the slab and its periodic images.
We first calculate the ARPES of the pure Bi$_2$Se$_3$ 2$\times$2$\times$9 slab, without any doping, and unfold
the band structures to the BZ of the 1$\times$1$\times$9 slab.
The results are shown in Fig.~\ref{fig:Bi2Se3}(a).
The unfolded spectra are identical to those calculated directly from  1$\times$1$\times$9 slab,
as expected. The band structure show a clear Dirac-cone like band dispersion, which is a good agreement with the ARPES. \cite{Xia2009,Chen2010}

We then calculate the ARPES of Tl-doped Bi$_2$Se$_3$ slab.
We randomly selected 7 of the 72 Bi atoms and replaced them with the Tl atoms,
which corresponds to about 10\% doping, in accordance with the experiments\cite{Chen2010}.
The lattice structures of the doped slabs are fully relaxed. The calculated ARPES
is shown in Fig.~\ref{fig:Bi2Se3}(b).
The ARPES are averaged from 8 randomly doped structures,
with contributions from both top and bottom sides of the slabs.
As we see, there are many surface states appear compared to the undoped Bi$_2$Se$_3$ slab.
However, most of the surface states are gaped, and thus are the trial states.
There is one band shows a notable gapless Dirac cone, with a strong bright spot at the Dirac point.
The calculated ARPES is a good agreement with the ARPES of Tl-doped Bi$_2$Se$_3$ at
the similar doping density \cite{Chen2010}.

Figure~\ref{fig:Bi2Se3}(c)depicts the calculated ARPES of the Fe-doped Bi$_2$Se$_3$.
Seven of the 72 Bi atoms in the slab are randomly selected and replaced by the Fe atoms.
The spins of the doped Fe atoms are aligned ferromagnetically.
The calculated ARPES is also averaged from the results of 8 random Fe doped slabs\cite{Chen2010}.
As shown in the figure, the surface states are gapped due to the breaking of the time reversal symmetry.

\subsubsection{ARPES for  Pb$_{1-x}$Sn$_{x}$Se}

\begin{figure*}
		\centering
		\includegraphics[width=0.6\textwidth]{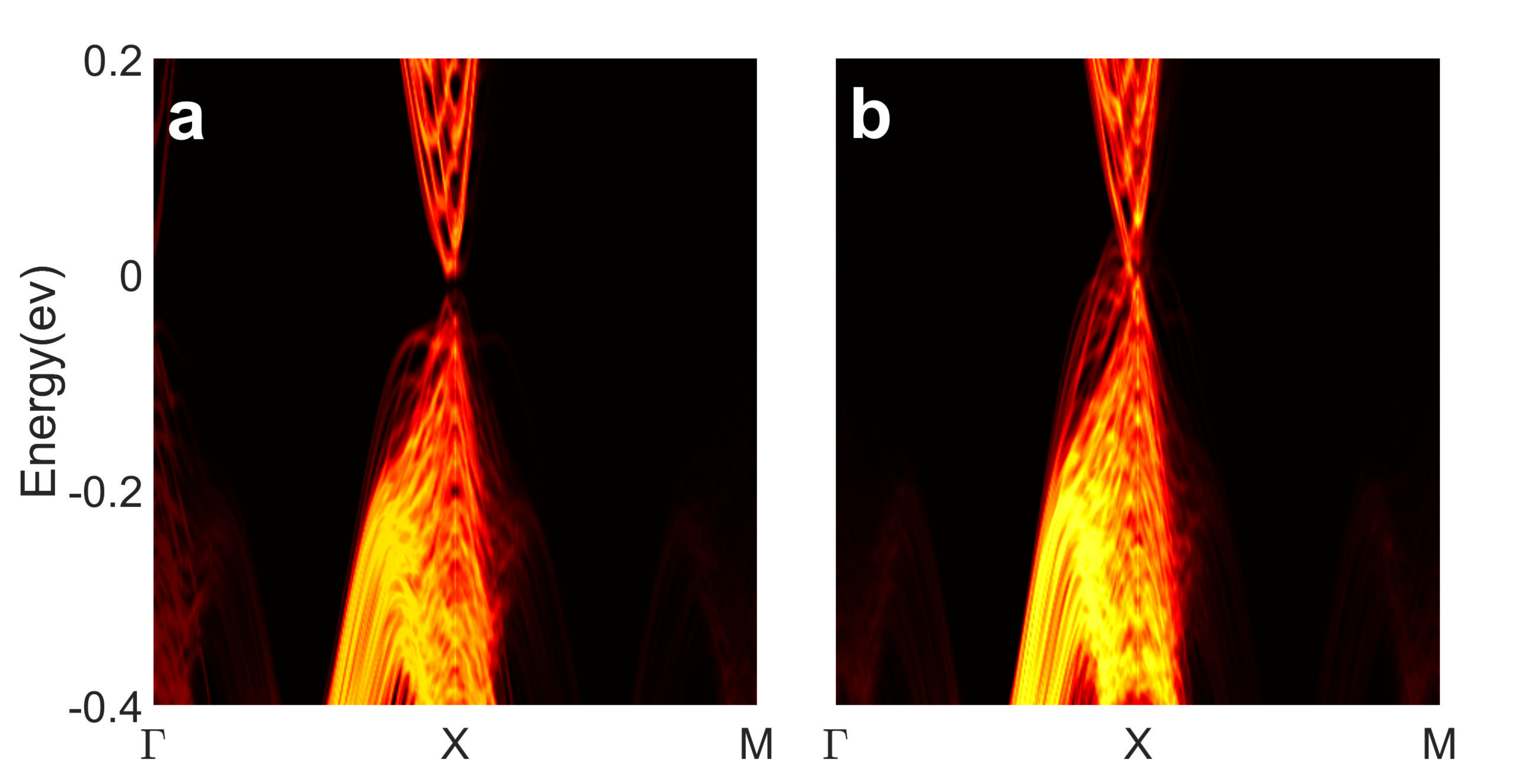}
		\caption{The ARPES spectra of Pb$_{1-x}$Sn$_{x}$Te with (a)$x$=0.2, and (b) $x$=0.4.}
		\label{fig:SnTe}
\end{figure*}

Shortly after the TI was discovered,  topological crystalline insulator (TCI) was predicted \cite{Fu2011} and discovered in SnTe\cite{Tanka2012},
Pb$_{1-x}$Sn$_{x}$Se \cite{Dziawa2012,Hsieh2012} and Pb$_{1-x}$Sn$_{x}$Te($x>0.3$) \cite{Xu2012}series compounds by the spin-resolved ARPES.
Similar to that of the TI, the TCIs also have the gapless surface states, which are protected by the crystalline symmetry.

We study the topological phase transition in the Pb$_{1-x}$Sn$_{x}$Te alloys.
At $x$=0 (i.e. pure PbSe) the system is a trivial insulator, whereas at $x$=1 (i.e. pure SnSe), the material is a TCI.
The band inversion transition, and thus the topological transition occurs around $x\simeq \frac{1}{3}$ \cite{Diammock1966} predicted by the theoretical calculations \cite{Gao2008}, and confirmed
by the ARPES experiments\cite{Xu2012}.
Here, we choose two representative compositions, namely $x$=0.2 and $x$=0.4, to investigate the change of surface states as function of $x$.
We construct a slab model containing 2$\times$2$\times$10  eight-atom PbTe convention cells. A 15 \AA ~vacuum is added to avoid the interactions between the slab and its periodic images.

For the $x$=0.2 structures, we randomly selected 32 of Pb atoms and replaced them with Sn atoms. The lattice constants are given by a linear
interpolation of lattice constants of SnTe and PbTe compounds, i.e.,
\begin{equation}
\boldsymbol{a}_{\rm alloy} \approx x \boldsymbol{a}_{\rm SnSe} +(1-x) \boldsymbol{a}_{\rm PbSe}\, .
\end{equation}
The atomic positions are then fully relaxed under the interpolated lattice constants.
The calculated ARPES of Pb$_{0.8}$Sn$_{0.2}$Te shown in Fig.~\ref{fig:SnTe}(a), which is also averaged from 8 random structures.
First-principles calculations show that Pb$_{0.8}$Sn$_{0.2}$Te is a trivial insulator\cite{Gao2008}.
Indeed, the calculated ARPES of the surface state have a obvious gap ($\sim$ 25 meV), which
is in a good agreement with the experimental results\cite{Hsieh2012,Tanka2012,Xu2012}.

	   For the Pb$_{0.6}$Sn$_{0.4}$Te alloys, the band inversion occurs at $L$ point, and the system becomes a TCI, as predicted
in \cite{Xu2012}, and the gapless Dirac point is predicted to  occurs along the mirror line ($\Gamma$-$X$-$\Gamma$) direction.
Figure~\ref{fig:SnTe}(b) depicts the calculated ARPES of Pb$_{0.6}$Sn$_{0.4}$Te, also averaged from 8 random structures.
As shown in the figure, the is a gapless surface state near the $X$ point, which is in a good agreement with the experimental ARPES \cite{Hsieh2012,Tanka2012,Xu2012}.
We note that the Dirac point of this system is not as bright as that in the TI systems.
Whereas the surface state of a TI is protected by the time reversal symmetry, which is very robust without
the magnetic impurities, the mirror symmetry in  Pb$_{0.6}$Sn$_{0.4}$Te alloys only exists in the sense of statistical averaged structures.

\section{Summary}

We develop an efficient band unfolding method based on the NAO bases, which can be used for the analysing  the band structures of
large and complex systems. We apply this method to investigate the surface states of the non-magnetically and magnetically doped topological
insulators Bi$_2$Se$_3$ and the topological crystalline insulators Pb$_{1-x}$Sn$_{x}$Te, and compare the results
with the ARPES experiments. The method provides a powerful tool to investigate the doping, alloying and disorder effects, which are
not limited to the topological materials.

\acknowledgements
This work was funded by the Chinese National Science
Foundation Grant Number 12134012. The numerical calculations were done on the USTC HPC facilities.

\appendix
\section{Expanding NAOs in the plane wave bases}

The key to our algorithm is to expand the NAOs in the plane wave bases, i.e.,
\begin{equation}
\phi_{\mu}(\boldsymbol{r})  = \int d\boldsymbol{q}\,  \phi_{\mu} (\boldsymbol{q})  e^{i\boldsymbol{q}\cdot \boldsymbol{r}} \, ,
\label{eq:phi_r}
\end{equation}
where,
\begin{equation}
\phi_{\mu} (\boldsymbol{q}) ={1\over \sqrt{V}} \int e^{-i \boldsymbol{q} \cdot \boldsymbol{
    r}} \phi_{\mu}  (\boldsymbol{r}) d \boldsymbol{r} \, .
    \label{eq:phi_q}
 \end{equation}
The NAOs  $\phi_{\mu}(\boldsymbol{r})$ are of the form of,
\begin{equation}
\phi_{\mu}(\boldsymbol{r}) = f_{\mu} (|\boldsymbol{r}|)Y_{lm}(\hat{r})\, ,
\end{equation}
where  $f_{\mu} (|\boldsymbol{r}|)$ is the redial function of the orbital, and  $Y_{lm}(\hat{r})$ is the spherical harmonic function.
To calculate Eq.~(\ref{eq:phi_q}), we use the relation,
\begin{equation}
e^{i\boldsymbol{q}\cdot \boldsymbol{ r}} = \sum_{l,m}
{4\pi(i^l)j_l(|\boldsymbol{q}||\boldsymbol{\bf r}|)Y_{lm}(\hat{q})Y_{lm}(\hat{r}) }\, ,
\label{eq:pw}
\end{equation}
where the $\hat{q}$ and $\hat{r}$ are the unit vectors of $\boldsymbol{q}$ and
$\boldsymbol{r}$. Plugging Eq.~(\ref{eq:pw}) into Eq.~(\ref{eq:phi_q}), we have
\begin{widetext}
\begin{equation}
\phi_{\mu} (\boldsymbol{q})
    = \sum_{l',m'} {4\pi(-i)^{l'} \int j_{l'}(|\boldsymbol{q}||{\bf r}|) r^2 f_\mu (r) dr
\int  Y_{lm}(\hat{r})Y_{l'm'}(\hat{r}) d\Omega \;\;
Y_{l'm'}(\hat{q}) }\, .
\end{equation}
\end{widetext}

Using the relation,
\begin{equation}
\int  Y_{lm}(\hat{r})Y_{l'm'}(\hat{r}) d\Omega = \delta_{l,l'} \delta_{m,m'}\, ,
\end{equation}
we obtain that,
\begin{equation}
\phi_{\mu}({\bf q}) =(-i)^l{4\pi\over \sqrt{V}}
\, Y_{lm}(\hat{q})
f_{\mu}(|{\bf q}|)\, ,
\end{equation}
where,
\begin{equation}
f_\mu(q) = \int_{0}^{\infty} { dr\, r^2 f_\mu(r)j_l(q r) }\, .
\end{equation}
In the code, we first calculate $f_\mu(q)$ for each type of the NAOs,
on the one-dimensional discrete $q_i$ meshes, and store them in a table.
The $f_\mu(q)$ at an arbitrary $q$ can be calculated via the Cubic spline interpolation.

\end{document}